\documentclass[aps,prd,preprint,12pt,superscriptaddress,nofootinbib,a4paper]{revtex4-1}

\usepackage[utf8]{inputenc}
\usepackage{amsmath,amssymb}
\usepackage[separate-uncertainty=true]{siunitx}
\usepackage{graphicx}
\usepackage[usenames,dvipsnames]{xcolor}
\usepackage[caption=false]{subfig}
\usepackage{hyperref}
\usepackage[capitalize]{cleveref}
\usepackage{booktabs}
\usepackage{tabularx}
\usepackage{xspace}
\usepackage{color}
\usepackage[normalem]{ulem}
\usepackage{slashed}
\usepackage{bbold}
\usepackage{wasysym}
\usepackage{graphicx}
\usepackage{mathrsfs} 

\allowdisplaybreaks

\graphicspath{{graphics/}}

\newcommand{\eqn}{equation}
\newcommand{\lb}{\left(}
\newcommand{\rb}{\right)}
\newcommand{\be}{\beta}
\newcommand{\al}{\alpha}

\newcommand{\GeV}{{\ensuremath\rm GeV}}

\newcommand{\pb}{{\ensuremath\rm pb}}
\newcommand{\fb}{{\ensuremath\rm fb}}

\DeclareSIUnit{\pb}{pb}
\DeclareSIUnit{\fb}{fb}
\AtBeginDocument{
\heavyrulewidth=.08em
\lightrulewidth=.05em
\cmidrulewidth=.03em
\belowrulesep=.65ex
\belowbottomsep=0pt
\aboverulesep=.4ex
\abovetopsep=0pt
\cmidrulesep=\doublerulesep
\cmidrulekern=.5em
\defaultaddspace=.5em

\newcolumntype{C}{>{\centering\arraybackslash}X}
\newcolumntype{b}{C}
\newcolumntype{s}{>{\hsize=.6\hsize}C}
\newcolumntype{R}{>{\raggedleft\arraybackslash}X}
}

\begin{document}
\bibliographystyle{hunsrt}
\date{\today}
\rightline{RBI-ThPhys-2024-18}
\title{{\Large Overview on low mass scalars at $e^+e^-$ facilities -
theory}}
\author{Tania Robens}
\affiliation{Ruder Boskovic Institute, Bijenicka cesta 54, 10000 Zagreb, Croatia}

\renewcommand{\abstractname}{\texorpdfstring{\vspace{0.5cm}}{} Abstract}

%%%%%%%%%%%%%%%%%%%%%%%%%%%%%%%
\begin{abstract}
    \vspace{0.5cm}
I give a short summary of scenarios with new physics scalars that could be investigated at future $e^+e^-$ colliders.  I concentrate on cases where at least one of the additional scalar has a mass below 125 GeV, and discuss models where this could be realized. In general, there are quite a few additional new physics scenarios and signatures that are still allowed by current constraints and should be investigated in more detail at future $e^+e^-$ machines. Most of the material presented here has already been discussed in \cite{Robens:2023bzp}, and I therefore try to focus on novel developments since then.
  
\end{abstract}
%%%%%%%%%%%%%%%%%%%%%%%%%%%%%%%

\maketitle

\section{Introduction}
In the European Strategy report \cite{EuropeanStrategyforParticlePhysicsPreparatoryGroup:2019qin,CERN-ESU-015}, Higgs factories were identified as one of the high priority projects after the HL-LHC. At such machines, the properties of the Higgs particles should be measurable to utmost precision. Furthermore, new physics scalar states could also be produced in the mass range up to $\sim\,160\,\GeV$ depending on the collider process.

In this short proceeding, I give a very brief overview on possible production channels and also give examples of models that can render low mass scalars. I here concentrate on novel developments since the LCWS 2023 proceedings \cite{Robens:2023bzp}. Additional material is also available in \cite{Robens:2022zgk}. Note that the material here is partly inspired by the discussion in the Focus Topics report of the ongoing ECFA study \cite{deBlas:2024bmz}.

\section{Processes at Higgs factories}

\begin{center}
\begin{figure}[htb!]
\begin{center}
%\begin{minipage}{0.5\textwidth}
\includegraphics[width=0.55\textwidth]{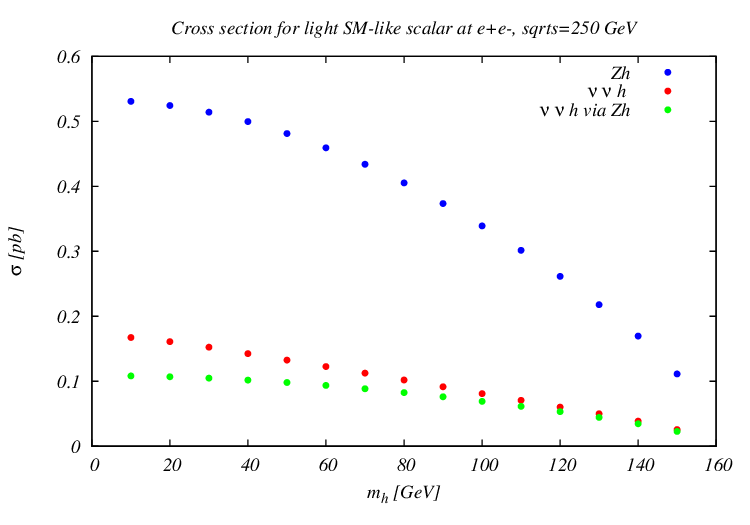}
\caption{\label{fig:prod250} {Leading-order production cross sections for $e^+\,e^-\,\rightarrow\,Z\,h$ {\sl (blue)} and $e^+\,e^-\,\rightarrow\,h\,\nu_\ell\,\bar{\nu}_\ell$ {\sl (red)} production at an $e^+\,e^-$ collider with a com energy of 250 \GeV~using Madgraph5 for an SM-like scalar $h$. We also display the contribution of $Z\,h$ to $\nu_\ell\,\bar{\nu}_\ell\,h$ using a factorized approach for the Z decay, generated via $e^+\,e^-\,\rightarrow\,Z\,h\,\times\,\text{BR}\lb h\,\rightarrow\,\nu_\ell\,\bar{\nu}_\ell \rb$ {\sl (green)}. Update of plot in \cite{Robens:2023bzp}.}}
%\end{minipage}
\end{center}
\end{figure}
\end{center}

At the center-of-mass (com) energies of Higgs factories, Higgs strahlung is the dominant production mode for single scalar production \cite{Abramowicz:2016zbo}. Leading-order predictions for $Zh$ production at $e^+e^-$ colliders for low mass scalars which are Standard Model (SM)-like, using Madgraph5 \cite{Alwall:2011uj}, are shown in figure \ref{fig:prod250} for a center-of-mass energy of 250 \GeV. The $e^+e^-\,\rightarrow\,h\,\nu_\ell\,\bar{\nu}_\ell$ process contains contributions from both scalar strahlung and VBF type topologies, so we also display the expected rates from the former for this final state using a factorized approach. It can be seen that for higher scalar masses the dominant contribution stems from $Z\,h$ production.

\section{Projections for additional searches and connections to electroweak phase transitions}

This section summarizes the findings already presented in \cite{Robens:2023bzp}. We just list these here as we consider them to be fundamental for the past and current status regarding searches at Higgs factories as well as possible connections to electroweak phase transitions.

\begin{center}
\begin{figure}[htb!]
\begin{center}
\includegraphics[width=0.4\textwidth, angle=-90]{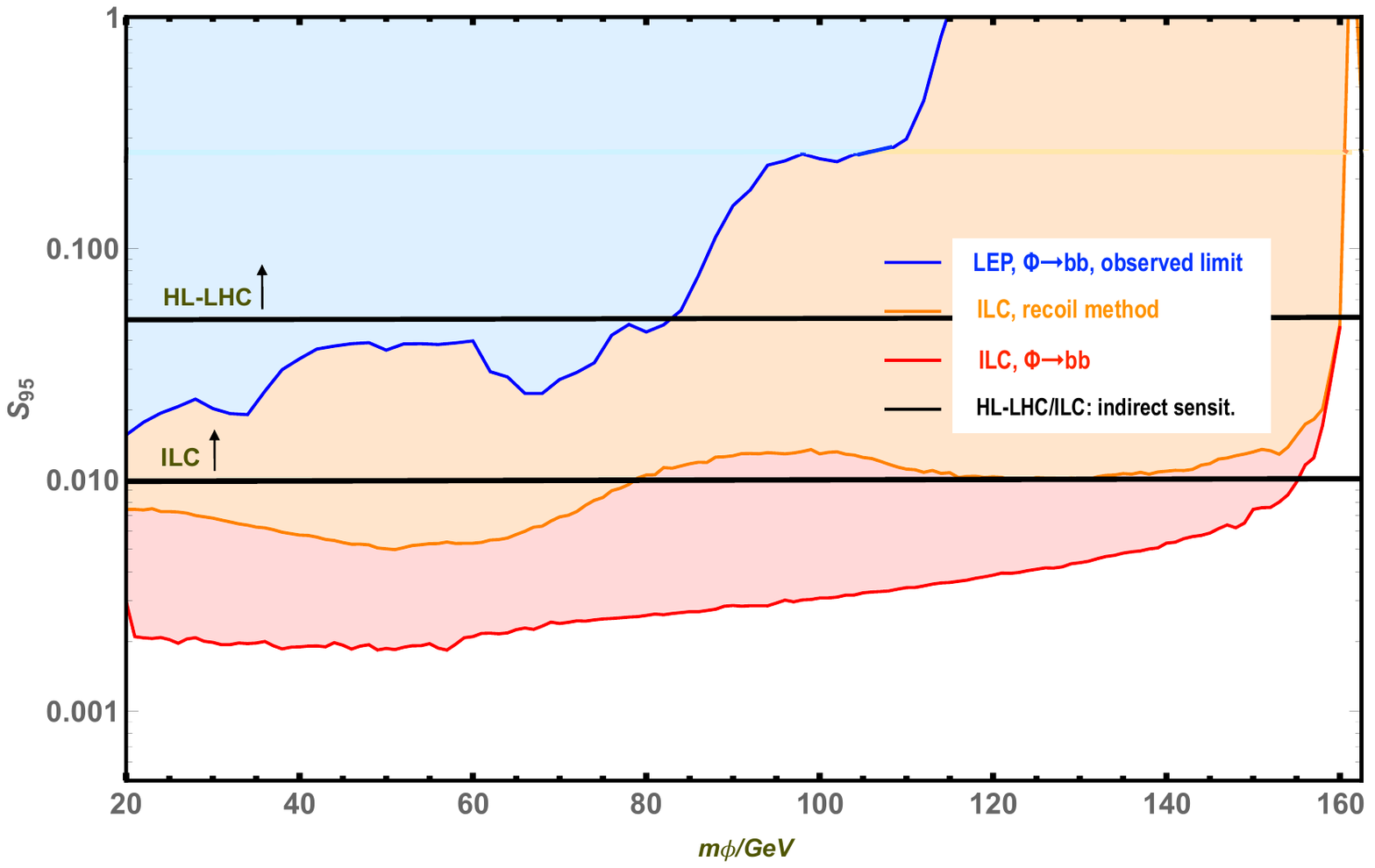}
\caption{\label{fig:lepgea} Sensitivity predictions for an ILC with a com energy of 250 \GeV~ from \cite{Drechsel:2018mgd}, as a function of the new scalars mass. The quantity $S_{95}$ is defined via Eqn. (\ref{eq:s95}). See text for further details.}
\end{center}
\end{figure}
\end{center}

A standard reference regarding the production of lighter scalars in scalar strahlung is given in \cite{Drechsel:2018mgd,Wang:2020lkq}. In principle two different analysis methods exist, which either use the pure $Z$ recoil ("recoil method") or take the light scalar decay into $b\,\bar{b}$ into account. We display the reach of these two methods, derived via
\begin{\eqn}\label{eq:s95}
S_{95}\,=\,\frac{\hat{\sigma}}{\sigma_\text{ref}}
\end{\eqn}
that gives an upper limit $\hat{\sigma}$ on a cross section compatible with the background only hypothesis at $95\%$ confidence level \cite{Drechsel:2018mgd}. The reference cross section $\sigma_\text{ref}$ is given by the expected production cross section of a SM-like scalar at the respective mass. This quantity can directly be translated into an upper bound on rescaling, see figure \ref{fig:lepgea}\footnote{Note that this figure was derived recasting results from LEP \cite{LEPWorkingGroupforHiggsbosonsearches:2003ing,ALEPH:2006tnd,OPAL:2002ifx}. Therefore, only the contribution from the $Z$ peak has been taken into account in the analysis. We refer the reader to the above references for a more detailed discussion regarding the treatment of these contributions.}.%. The figure is taken from \cite{Drechsel:2018mgd}. 
%The authors validate their method by reproducing LEP results \cite{LEPWorkingGroupforHiggsbosonsearches:2003ing,ALEPH:2006tnd} for these channels prior to applying their method to the ILC. 
%The results are shown in figure \ref{fig:lepgea}.

\begin{center}
\begin{figure}[htb!]
\begin{center}
\includegraphics[width=0.45\textwidth]{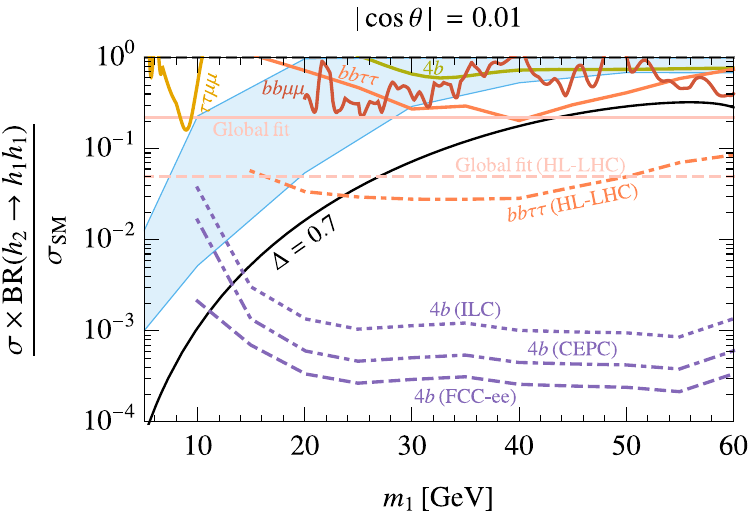}
\includegraphics[width=0.45\textwidth]{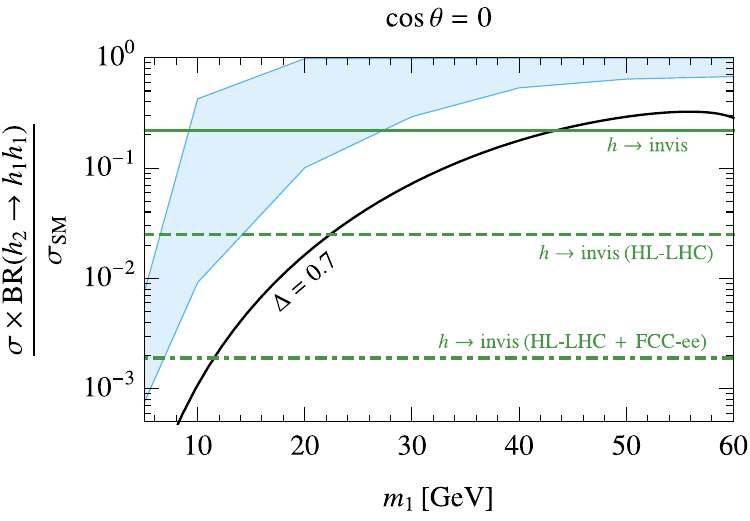}
\end{center}
\caption{\label{fig:ewps} Expected bounds on Higgs production via Higgs strahlung and subsequent decay into two light scalars, in the singlet extension scenario discussed in \cite{Kozaczuk:2019pet,Wang:2022dkz}. For $\cos\theta\,=\,0$ the constraints mainly stem from $h_{125}\,\rightarrow\,\text{invisible}$ searches. The blue shaded area corresponds to the region that allows for a strong first order electroweak phase transition in agreement with the scans performed in \cite{Kozaczuk:2019pet}. $\Delta R$ is a measure of the relative difference of realized singlet vacua to the true vacuum over the difference of the maximal barrier field configuration again with respect to the true vacuum (see text for details). Depending on $m_1$ this scenario can be tested at current or future collider experiments.}
\end{figure}
\end{center}

Another important topic is the connection of models with extended scalar sectors with different scenarios of electroweak phase transitions. In particular, for scenarios where the second scalar is lighter than a SM like candidate, such states can be investigated in Higgs-strahlung. A tell-tale signature in this case is the associated decay of $h\,\rightarrow\,h_i\,h_i$. The clean environment of a lepton collider typically allows to test the respective parameter space into regions that provide a priori relatively low rates. There has a been a lot if recent activity in this field; in figure \ref{fig:ewps}, we here exemplarily show results from \cite{Kozaczuk:2019pet}. This study assumes a simple singlet extension where the SM scalar sector is augmented by an additional scalar. The mixing between gauge and mass eigenstates of the SM-like doublet and the singlet is denoted by $\cos\theta$, where $\cos\theta\,=\,0$ denotes the SM decoupling scenario of no mixing. Several collider sensitivity projections are shown, including generic bounds for various discalar decay modes that have originally been derived in \cite{Liu:2016zki}. Show is also the region where strong first order electroweak phase transitions are possible following the analysis in \cite{Kozaczuk:2019pet} {\sl (light blue band)}, as well as a lower bound on the respective normalized cross sections guaranteeing a successful completion of the phase transition, denoted by $\Delta R\,=\,0.7$\footnote{The quantity $\Delta R\,=\,\frac{V\lb \phi_s\,T^*\rb-V\lb \phi_h,T^* \rb}{V\lb \phi_b\,T^*\rb-V\lb \phi_h,T^* \rb}$ is taken as a measure of the so-called thin-wall limit \cite{Kozaczuk:2019pet}, where it was numerically found that $\Delta R\,\geq\,0.7$ seems to provide a sufficient condition to evade this region. Here, $T^*$ denotes the temperature at which the transition is completed, and $\phi_s,\,\phi_h,$ and $\phi_b$ denote the singlet, broken phase, and location of barrier peak along the tunneling trajectory, respectively.}. It is evident that $e^+e^-$ Higgs factories would be an ideal environment to confirm or rule out such scenarios.

\section{Parameter space for some sample models}

After briefly discussing new physics signatures, we now turn to models that still allow for such low mass scalars. This is obviously only a brief overview, and more models might exist allong for low mass scalars accessible at Higgs factories; see e.g. \cite{Robens:2022zgk} for more details.

\begin{center}
\begin{figure}[htb!]
\begin{center}
\includegraphics[width=0.48\textwidth]{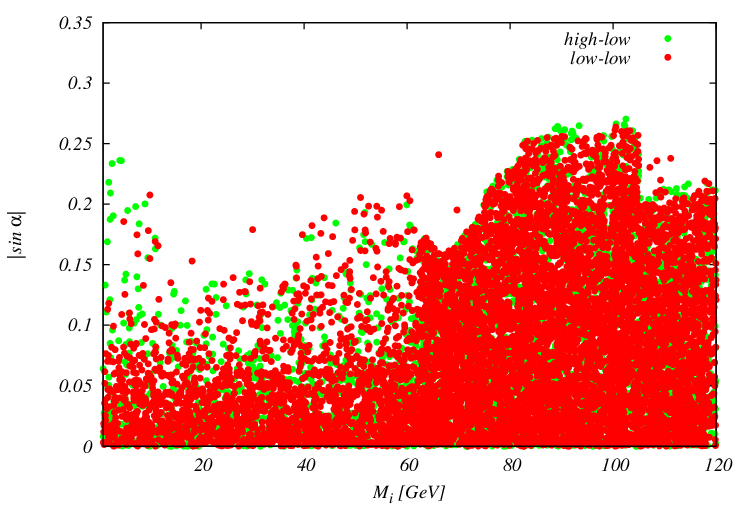}
\includegraphics[width=0.48\textwidth]{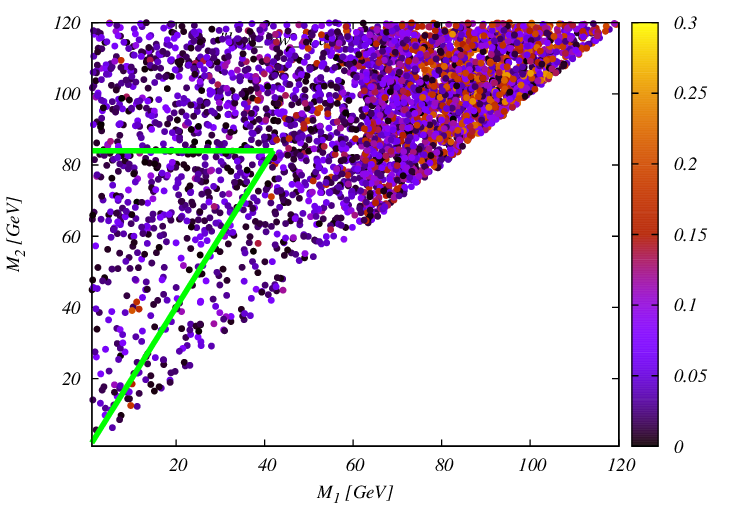}
\caption{\label{fig:trsm} {Available parameter space in the TRSM, with one (high-low) or two (low-low) masses lighter than 125 \GeV. {\sl Left}: light scalar mass and mixing angle, with $\sin\al\,=\,0$ corresponding to complete decoupling. {\sl Right:} available parameter space in the $\lb m_{h_1},\,m_{h_2}\rb$ plane, with color coding denoting the rescaling parameter $\sin\al$ for the lighter scalar $h_1$. Within the green triangle, $h_{125}\,\rightarrow\,h_2 h_1\,\rightarrow\,h_1\,h_1\,h_1$ decays are kinematically allowed. Taken from \cite{Robens:2022zgk}.}}
\end{center}
\end{figure}
\end{center}

The first model we discuss is a model that extends the scalar sector of the SM by two additional fields that transform as singlets under the electroweak gauge group, the Two Real Singlet Model (TRSM) \cite{Robens:2019kga,Robens:2022nnw}. This model contains three CP-even neutral scalars that relate the gauge and mass eigenstates $h_{1,2,3}$ via mixing. One of the scalars necessarily needs to comply with current LHC findings for the SM-like 125 \GeV resonance. However, the other two can in principle take any mass values, as long as all current theoretical and experimental constraints are fulfilled.
A detailed discussion including all constraints can be found in \cite{Robens:2019kga,Robens:2022nnw}, with recent updates on benchmark planes also presented in \cite{Robens:2023oyz}. In figure \ref{fig:trsm}, two cases are shown where either one (high-low) or two (low-low) scalar masses are smaller than $125\,\GeV$. On the y-axis, the respective mixing angle is shown. Decoupling here corresponds to $\sin\al\,=\,0$. The parameter space is basically identical to the one discussed in \cite{Robens:2023bzp}. We use the scan discussed in \cite{Robens:2019kga,Robens:2022nnw}, making use of the ScannerS framework \cite{Coimbra:2013qq,Muhlleitner:2020wwk}. Direct search constraint are implemented via HiggsTools \cite{Bahl:2022igd} using the most up to date publicly available version. This codes includes HiggsBounds \cite{Bechtle:2008jh,Bechtle:2011sb,Bechtle:2013wla,Bechtle:2020pkv} and HiggsSignals \cite{Bechtle:2013xfa,Bechtle:2014ewa,Bechtle:2020uwn} for testing direct searches and Higgs signal strength constraints, respectively.

Another interesting scenario is given by two Higgs doublet models, where the SM scalar sector is augmented by a second doublet. Such models have been studied by many authors, see e.g. \cite{Branco:2011iw} for an overview on the associated phenomenology. We here focus on the available parameter space in 2HDMs of type I, where all leptons couple to one doublet only. In such models, flavour constraints that are severly constraining the parameter space for other Yukawa structures (see e.g. \cite{Misiak:2017bgg} for a discussion in type II models) are leviated.

\begin{center}
\begin{figure}
\begin{center}
\includegraphics[width=0.35\textwidth]{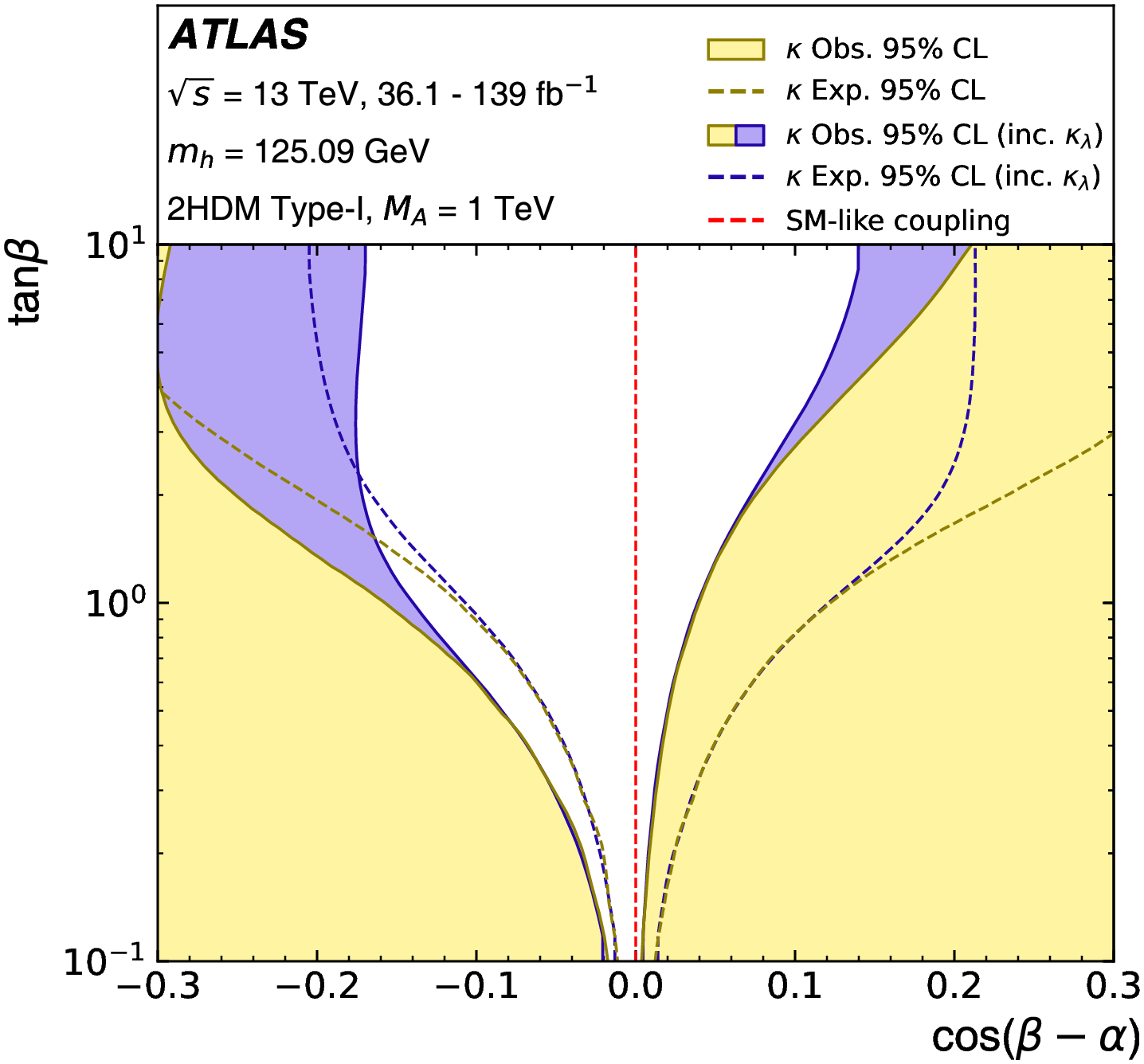}
\includegraphics[width=0.35\textwidth]{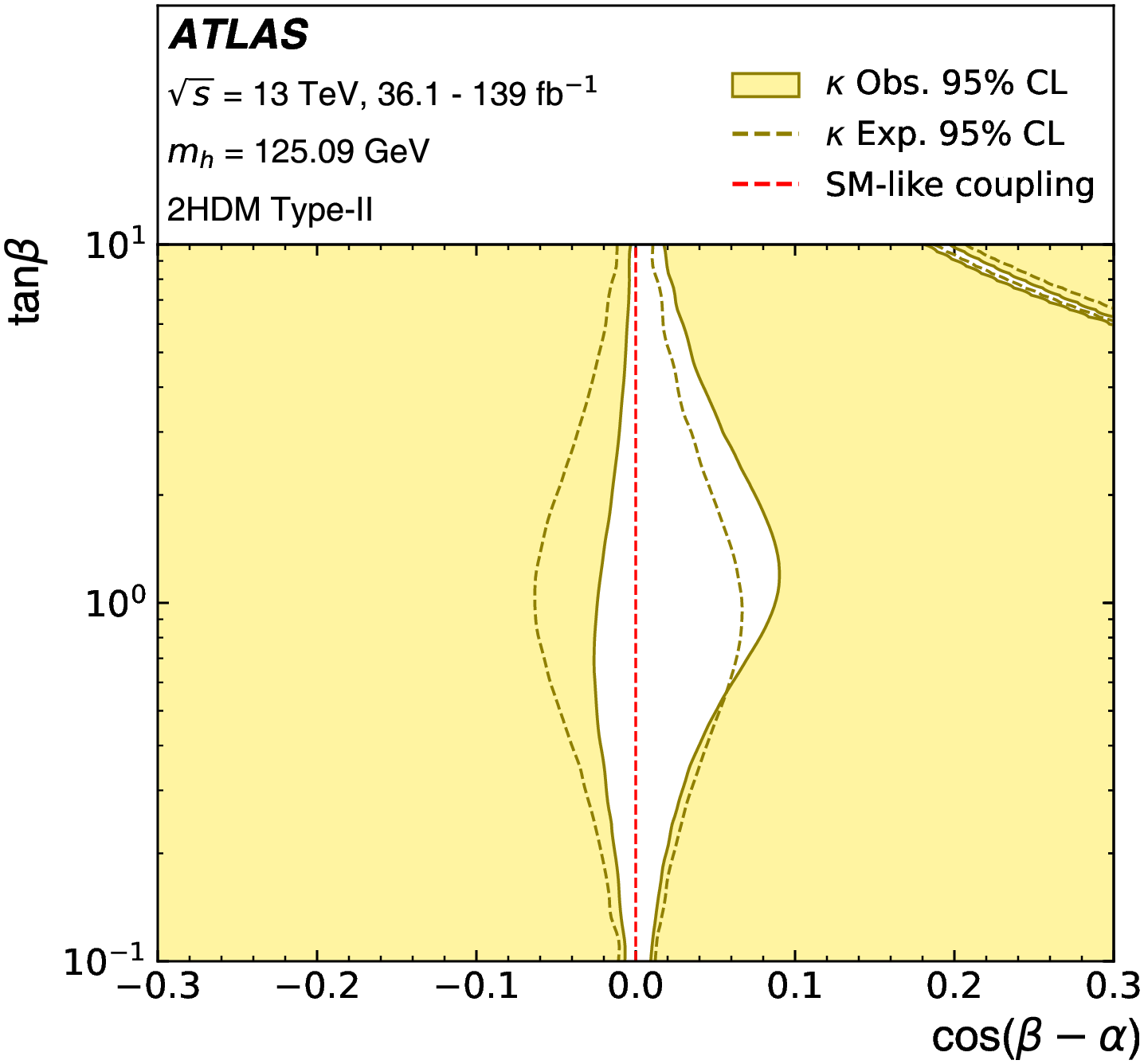}\\
\vspace{-2mm}
\includegraphics[width=0.35\textwidth]{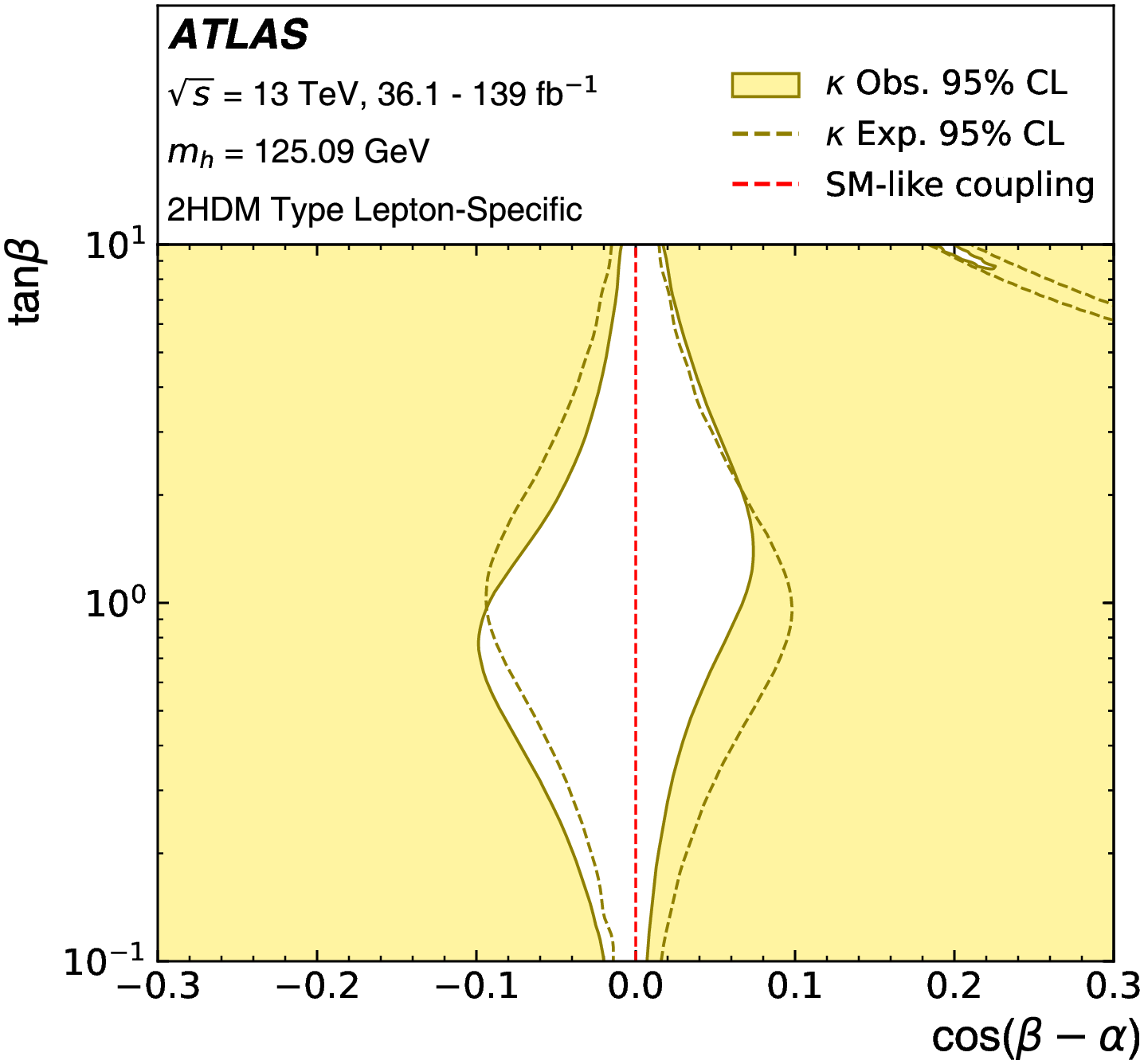}
\includegraphics[width=0.35\textwidth]{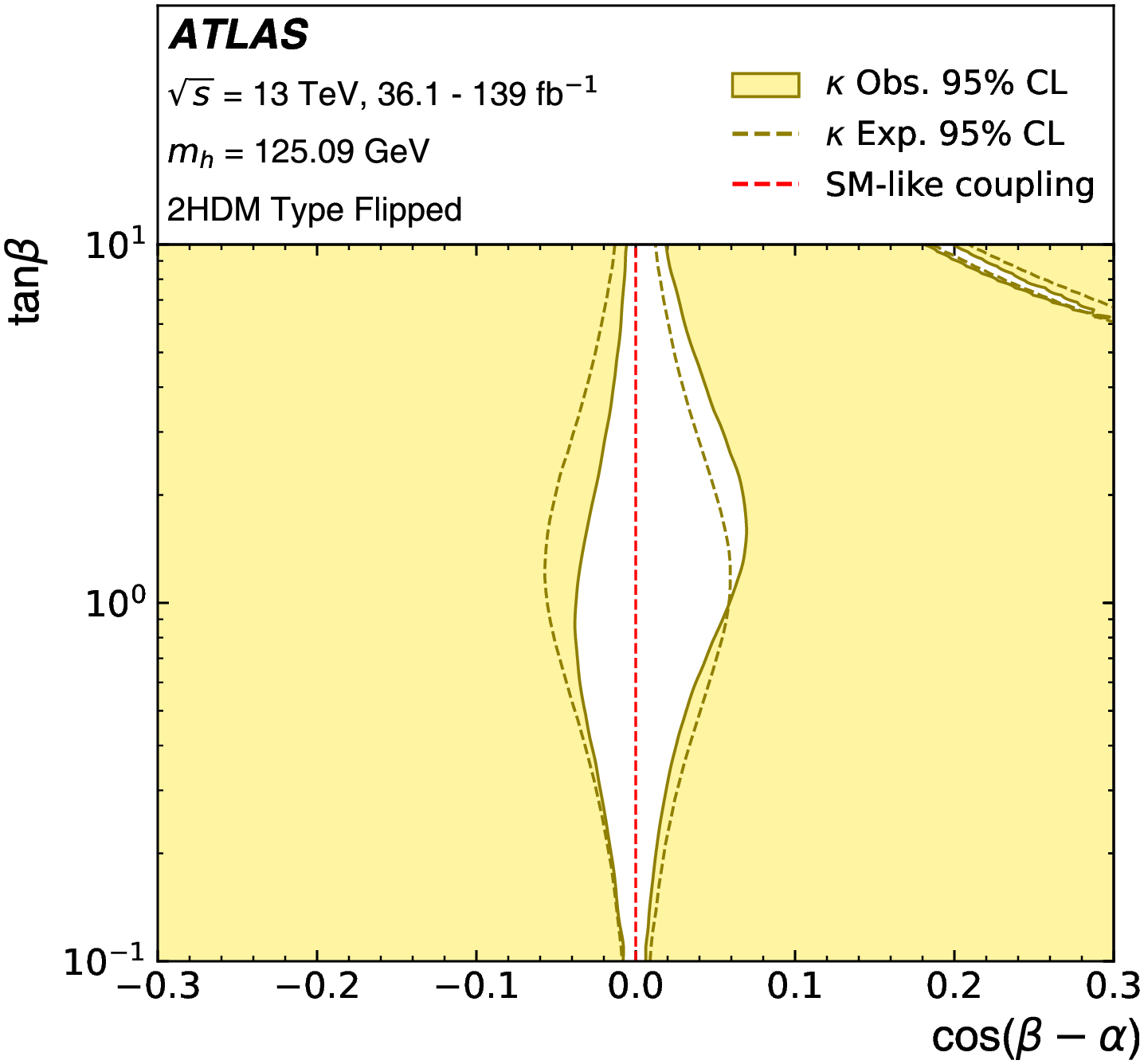}
\end{center}
\caption{\label{fig:atldec} Allowed and excluded regions in the $\lb \cos\lb \be-\al\rb,\tan\be \rb$ parameter space from Higgs signal strength from the ATLAS collaboration \cite{ATLAS:2024lyh}, for various Yukawa structures.}
\end{figure}
\end{center}

Apart from the masses of the additional particles, important parameters in 2HDMs are given by a combination of the different mixing angles from gauge to mass eigentstates, $\cos\lb \be-\al \rb$, as well as $\tan\be$ that describes the ratio of the vacuum expectation values in a certain basis\footnote{See e.g. \cite{Davidson:2005cw} for a discussion of the meaning of physical parameters in a basis independent way.}. In general, if the lighter neutral CP-even scalar $h$ is identified with the 125 \GeV~ resonance, the decoupling limit $\cos\lb \be - \al\rb\,=\,0$ denotes the configuration where the couplings to vector boson become SM-like. The Higgs signal strength poses important constraints on this quantity. In figure \ref{fig:atldec}, we display the current constraints from the ATLAS collaboration using the 125 \GeV scalar signal strength for various types of Yukawa structures \cite{ATLAS:2024lyh}.

Another question of interest is which parameter space is still allowed in such models after all possible other theoretical and experimental constraints are taken into account. For this, we make use of the publicly available tool thdmTools  \cite{Biekotter:2023eil}, that allows for efficient scans of the 2HDM parameter space and contains an internal interface to HiggsTools for comparison with null-results from collider searches.

As the minimal version of the 2HDM contains in total 6 free parameters after one of the CP even neutral scalar masses has been fixed to guarantee agreement with current LHC findings, it is customary to reduce the number of these parameters in typical scan plots to 2 dimensional planes. In such scenarios, care must be taken to not misinterpret the results; often, limits can change significantly once these underlying assumptions about parameter relations are altered.

We here exemplarily show the parameter space for a type I 2HDM, where we fixed all additional scalar masses to the same value $m_H\,=\,m_A\,=\,m_{H^\pm}$ and additionally set $\cos\lb \be-\al\rb$ to a certain value. In one of the two scenarios shown here, we additionally fixed the additional potential parameter $m_{12}^2\,=\,m_H^2\,\sin\be\,\cos\be$.

\begin{center}
\begin{figure}
\begin{center}
\includegraphics[width=0.49\textwidth]{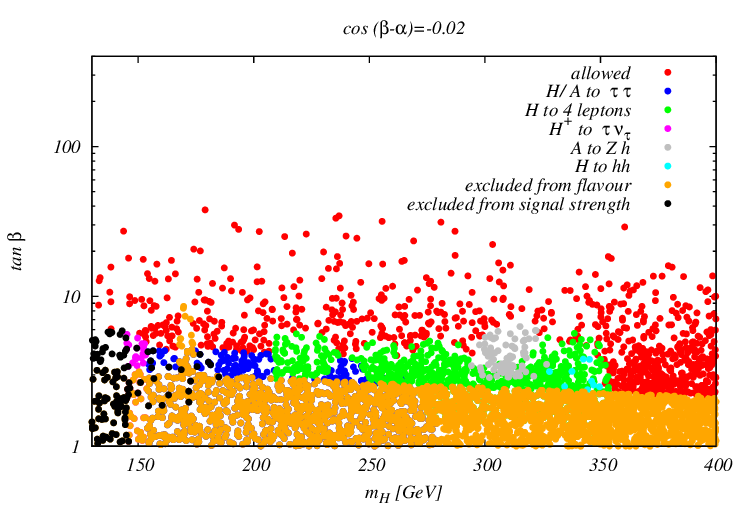}
\includegraphics[width=0.49\textwidth]{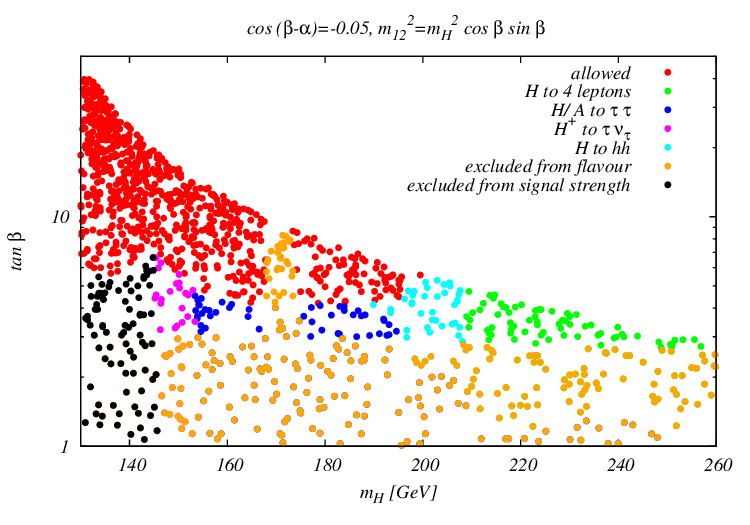}
\caption{\label{fig:2hdm} Allowed regions in the $\lb m_H,\,\tan\be\rb$ plane after various constraints have been taken into account, including experimental searches at the LHC as currently available using thdmTools. {\sl Left:} Here, $\cos\lb \be-\al\rb\,=\,-0.02$, and $m_{12}^2$ is floating freely. {\sl Right:} We here fixed $\cos\lb \be-\al\rb\,=\,-0.05,\,m_{12}^2\,=\,m_H^2\,\sin\be\,\cos\be$. See text for further details.}
\end{center}
\end{figure}
\end{center}

We show the corresponding parameter space in figure \ref{fig:2hdm}. Note that we chose to require a set of initial conditions, such as vacuum stability, perurtbative unitarity, and agreement with electroweak precision data to be fulfilled. Shown are then the exclusion bounds from flavour physics (via an internal interface to SuperIso \cite{Mahmoudi:2007vz,Mahmoudi:2008tp}), Higgs signal strength, as well as currently implemented collider searches. The relevant searches in this case are searches looking for a heavy resonance decaying into 4 lepton final states via electroweak gauge bosons \cite{ATLAS:2020tlo}, heavy resonances decaying into di-tau final states \cite{CMS:2022goy}, searches for charged scalars into leptonic decay modes \cite{ATLAS:2018gfm}, resonance-enhanced discalar searches \cite{CMS:2018amk}, and the process $A\,\rightarrow\,h\,Z$ \cite{CMS:2019qcx}. 

In general, for the two scenarios shown here, the available parameter space is already severely constrained from current searches at the LHC. For the scenario where $\cos\lb \be -\al\rb=-0.02$, current searches cease to be sensitive for masses $\gtrsim\,400\,\GeV$, but we can find allowed parameter points in the whole mass range, where a lower scale for $\tan\be$ is set by flavour constraints. On the other hand, if we additionally fix the potential parameter $m_{12}^2$, we see that there is also an upper limit on $\tan\be$ from the remaining constraints, basically leading to a maximal mass scale of around 200 \GeV~ for the heavy scalars. Note that these bounds only apply when the other free parameters are fixed as discussed.

\section{Conclusions}
I very briefly discussed some aspects of searches for low mass scalars at Higgs factories, including models that allo for such low mass states, and provided references for further reading. In particular, novel studies exceeding the ones presented here are highly encouraged and could be included as an input for e.g. the next European Strategy update.
\section{Acknowledgements}
TR acknowledges financial support from the Croatian Science Foundation (HRZZ) project " Beyond the Standard Model discovery and Standard Model precision at LHC Run III", IP-2022-10-2520. I also thank K. Radchenko, T. Biekotter and H. Bahl for continuous support for thdmTools and HiggsTools/ HiggsBounds/ HiggsSignals.
\bibliography{lit,lit_lw,lit_epi}

\begin{thebibliography}{10}

\bibitem{Robens:2023bzp}
Tania Robens.
\newblock {A short overview on low mass scalars at future lepton colliders}.
\newblock In {\em {International Workshop on Future Linear Colliders}}, 7 2023,
  2307.15962.

\bibitem{EuropeanStrategyforParticlePhysicsPreparatoryGroup:2019qin}
Richard~Keith Ellis et~al.
\newblock {Physics Briefing Book}: {Input for the European Strategy for
  Particle Physics Update 2020}.
\newblock 10 2019, 1910.11775.

\bibitem{CERN-ESU-015}
{2020 Update of the European Strategy for Particle Physics (Brochure)}.
\newblock Technical report, Geneva, 2020.

\bibitem{Robens:2022zgk}
Tania Robens.
\newblock {A Short Overview on Low Mass Scalars at Future Lepton Colliders}.
\newblock {\em Universe}, 8:286, 2022, 2205.09687.

\bibitem{deBlas:2024bmz}
Jorge de~Blas et~al.
\newblock {Focus topics for the ECFA study on Higgs / Top / EW factories}.
\newblock 1 2024, 2401.07564.

\bibitem{Abramowicz:2016zbo}
H.~Abramowicz et~al.
\newblock {Higgs physics at the CLIC electron\textendash{}positron linear
  collider}.
\newblock {\em Eur. Phys. J. C}, 77(7):475, 2017, 1608.07538.

\bibitem{Alwall:2011uj}
Johan Alwall, Michel Herquet, Fabio Maltoni, Olivier Mattelaer, and Tim
  Stelzer.
\newblock {MadGraph 5 : Going Beyond}.
\newblock {\em JHEP}, 06:128, 2011, 1106.0522.

\bibitem{Drechsel:2018mgd}
P.~Drechsel, G.~Moortgat-Pick, and G.~Weiglein.
\newblock {Prospects for direct searches for light Higgs bosons at the ILC with
  250 GeV}.
\newblock {\em Eur. Phys. J. C}, 80(10):922, 2020, 1801.09662.

\bibitem{Wang:2020lkq}
Yan Wang, Mikael Berggren, and Jenny List.
\newblock {ILD Benchmark: Search for Extra Scalars Produced in Association with
  a $Z$ boson at $\sqrt{s}=500$ GeV}.
\newblock 5 2020, 2005.06265.

\bibitem{LEPWorkingGroupforHiggsbosonsearches:2003ing}
R.~Barate et~al.
\newblock {Search for the standard model Higgs boson at LEP}.
\newblock {\em Phys. Lett. B}, 565:61--75, 2003, hep-ex/0306033.

\bibitem{ALEPH:2006tnd}
S.~Schael et~al.
\newblock {Search for neutral MSSM Higgs bosons at LEP}.
\newblock {\em Eur. Phys. J. C}, 47:547--587, 2006, hep-ex/0602042.

\bibitem{OPAL:2002ifx}
G.~Abbiendi et~al.
\newblock {Decay mode independent searches for new scalar bosons with the OPAL
  detector at LEP}.
\newblock {\em Eur. Phys. J. C}, 27:311--329, 2003, hep-ex/0206022.

\bibitem{Kozaczuk:2019pet}
Jonathan Kozaczuk, Michael~J. Ramsey-Musolf, and Jessie Shelton.
\newblock {Exotic Higgs boson decays and the electroweak phase transition}.
\newblock {\em Phys. Rev. D}, 101(11):115035, 2020, 1911.10210.

\bibitem{Liu:2016zki}
Zhen Liu, Lian-Tao Wang, and Hao Zhang.
\newblock {Exotic decays of the 125 GeV Higgs boson at future $e^+e^-$ lepton
  colliders}.
\newblock {\em Chin. Phys. C}, 41(6):063102, 2017, 1612.09284.

\bibitem{Wang:2022dkz}
Zhen Wang, Xuliang Zhu, Elham~E. Khoda, Shih-Chieh Hsu, Nikolaos
  Konstantinidis, Ke~Li, Shu Li, Michael~J. Ramsey-Musolf, Yanda Wu, and
  Yuwen~E. Zhang.
\newblock {Study of Electroweak Phase Transition in Exotic Higgs Decays at the
  CEPC}.
\newblock In {\em {Snowmass 2021}}, 3 2022, 2203.10184.

\bibitem{Robens:2019kga}
Tania Robens, Tim Stefaniak, and Jonas Wittbrodt.
\newblock {Two-real-scalar-singlet extension of the SM: LHC phenomenology and
  benchmark scenarios}.
\newblock {\em Eur. Phys. J. C}, 80(2):151, 2020, 1908.08554.

\bibitem{Robens:2022nnw}
Tania Robens.
\newblock {Two-Real-Singlet-Model Benchmark Planes}.
\newblock {\em Symmetry}, 15(1):27, 2023, 2209.10996.

\bibitem{Robens:2023oyz}
Tania Robens.
\newblock {TRSM benchmark planes - EPS-HEP2023 update}.
\newblock {\em PoS}, EPS-HEP2023:055, 2024, 2310.18045.

\bibitem{Coimbra:2013qq}
Rita Coimbra, Marco O.~P. Sampaio, and Rui Santos.
\newblock {ScannerS: Constraining the phase diagram of a complex scalar singlet
  at the LHC}.
\newblock {\em Eur. Phys. J. C}, 73:2428, 2013, 1301.2599.

\bibitem{Muhlleitner:2020wwk}
Margarete M\"uhlleitner, Marco O.~P. Sampaio, Rui Santos, and Jonas Wittbrodt.
\newblock {ScannerS: parameter scans in extended scalar sectors}.
\newblock {\em Eur. Phys. J. C}, 82(3):198, 2022, 2007.02985.

\bibitem{Bahl:2022igd}
Henning Bahl, Thomas Biek\"otter, Sven Heinemeyer, Cheng Li, Steven Paasch,
  Georg Weiglein, and Jonas Wittbrodt.
\newblock {HiggsTools: BSM scalar phenomenology with new versions of
  HiggsBounds and HiggsSignals}.
\newblock {\em Comput. Phys. Commun.}, 291:108803, 2023, 2210.09332.

\bibitem{Bechtle:2008jh}
Philip Bechtle, Oliver Brein, Sven Heinemeyer, Georg Weiglein, and Karina~E.
  Williams.
\newblock {HiggsBounds: Confronting Arbitrary Higgs Sectors with Exclusion
  Bounds from LEP and the Tevatron}.
\newblock {\em Comput. Phys. Commun.}, 181:138--167, 2010, 0811.4169.

\bibitem{Bechtle:2011sb}
Philip Bechtle, Oliver Brein, Sven Heinemeyer, Georg Weiglein, and Karina~E.
  Williams.
\newblock {HiggsBounds 2.0.0: Confronting Neutral and Charged Higgs Sector
  Predictions with Exclusion Bounds from LEP and the Tevatron}.
\newblock {\em Comput. Phys. Commun.}, 182:2605--2631, 2011, 1102.1898.

\bibitem{Bechtle:2013wla}
Philip Bechtle, Oliver Brein, Sven Heinemeyer, Oscar St\r{a}l, Tim Stefaniak,
  Georg Weiglein, and Karina~E. Williams.
\newblock {$\mathsf{HiggsBounds}-4$: Improved Tests of Extended Higgs Sectors
  against Exclusion Bounds from LEP, the Tevatron and the LHC}.
\newblock {\em Eur. Phys. J. C}, 74(3):2693, 2014, 1311.0055.

\bibitem{Bechtle:2020pkv}
Philip Bechtle, Daniel Dercks, Sven Heinemeyer, Tobias Klingl, Tim Stefaniak,
  Georg Weiglein, and Jonas Wittbrodt.
\newblock {HiggsBounds-5: Testing Higgs Sectors in the LHC 13 TeV Era}.
\newblock {\em Eur. Phys. J. C}, 80(12):1211, 2020, 2006.06007.

\bibitem{Bechtle:2013xfa}
Philip Bechtle, Sven Heinemeyer, Oscar St\r{a}l, Tim Stefaniak, and Georg
  Weiglein.
\newblock {$HiggsSignals$: Confronting arbitrary Higgs sectors with
  measurements at the Tevatron and the LHC}.
\newblock {\em Eur. Phys. J. C}, 74(2):2711, 2014, 1305.1933.

\bibitem{Bechtle:2014ewa}
Philip Bechtle, Sven Heinemeyer, Oscar St\r{a}l, Tim Stefaniak, and Georg
  Weiglein.
\newblock {Probing the Standard Model with Higgs signal rates from the
  Tevatron, the LHC and a future ILC}.
\newblock {\em JHEP}, 11:039, 2014, 1403.1582.

\bibitem{Bechtle:2020uwn}
Philip Bechtle, Sven Heinemeyer, Tobias Klingl, Tim Stefaniak, Georg Weiglein,
  and Jonas Wittbrodt.
\newblock {HiggsSignals-2: Probing new physics with precision Higgs
  measurements in the LHC 13 TeV era}.
\newblock {\em Eur. Phys. J. C}, 81(2):145, 2021, 2012.09197.

\bibitem{Branco:2011iw}
G.~C. Branco, P.~M. Ferreira, L.~Lavoura, M.~N. Rebelo, Marc Sher, and Joao~P.
  Silva.
\newblock {Theory and phenomenology of two-Higgs-doublet models}.
\newblock {\em Phys. Rept.}, 516:1--102, 2012, 1106.0034.

\bibitem{Misiak:2017bgg}
Mikolaj Misiak and Matthias Steinhauser.
\newblock {Weak radiative decays of the B meson and bounds on $M_{H^\pm }$ in
  the Two-Higgs-Doublet Model}.
\newblock {\em Eur. Phys. J. C}, 77(3):201, 2017, 1702.04571.

\bibitem{Davidson:2005cw}
Sacha Davidson and Howard~E. Haber.
\newblock {Basis-independent methods for the two-Higgs-doublet model}.
\newblock {\em Phys. Rev. D}, 72:035004, 2005, hep-ph/0504050.
\newblock [Erratum: Phys.Rev.D 72, 099902 (2005)].

\bibitem{ATLAS:2024lyh}
Georges Aad et~al.
\newblock {Interpretations of the ATLAS measurements of Higgs boson production
  and decay rates and differential cross-sections in $pp$ collisions at
  $\sqrt{s}=13$ TeV}.
\newblock 2 2024, 2402.05742.

\bibitem{Biekotter:2023eil}
Thomas Biek\"otter, Sven Heinemeyer, Jose~Miguel No, Kateryna Radchenko,
  Mar\'\i{}a Olalla~Olea Romacho, and Georg Weiglein.
\newblock {First shot of the smoking gun: probing the electroweak phase
  transition in the 2HDM with novel searches for A $\rightarrow$ ZH in $
  {\ell}^{+}{\ell}^{-}t\overline{t} $ and $ \nu \nu b\overline{b} $ final
  states}.
\newblock {\em JHEP}, 01:107, 2024, 2309.17431.

\bibitem{Mahmoudi:2007vz}
F.~Mahmoudi.
\newblock {SuperIso: A Program for calculating the isospin asymmetry of B
  ---\ensuremath{>} K* gamma in the MSSM}.
\newblock {\em Comput. Phys. Commun.}, 178:745--754, 2008, 0710.2067.

\bibitem{Mahmoudi:2008tp}
F.~Mahmoudi.
\newblock {SuperIso v2.3: A Program for calculating flavor physics observables
  in Supersymmetry}.
\newblock {\em Comput. Phys. Commun.}, 180:1579--1613, 2009, 0808.3144.

\bibitem{ATLAS:2020tlo}
Georges Aad et~al.
\newblock {Search for heavy resonances decaying into a pair of Z bosons in the
  $\ell ^+\ell ^-\ell '^+\ell '^-$ and $\ell ^+\ell ^-\nu {{\bar{\nu }}}$ final
  states using 139 $\mathrm {fb}^{-1}$ of proton\textendash{}proton collisions
  at $\sqrt{s} = 13\,$TeV with the ATLAS detector}.
\newblock {\em Eur. Phys. J. C}, 81(4):332, 2021, 2009.14791.

\bibitem{CMS:2022goy}
Armen Tumasyan et~al.
\newblock {Searches for additional Higgs bosons and for vector leptoquarks in
  $\tau\tau$ final states in proton-proton collisions at $\sqrt{s}$ = 13 TeV}.
\newblock {\em JHEP}, 07:073, 2023, 2208.02717.

\bibitem{ATLAS:2018gfm}
Morad Aaboud et~al.
\newblock {Search for charged Higgs bosons decaying via $H^{\pm} \to
  \tau^{\pm}\nu_{\tau}$ in the $\tau$+jets and $\tau$+lepton final states with
  36 fb$^{-1}$ of $pp$ collision data recorded at $\sqrt{s} = 13$ TeV with the
  ATLAS experiment}.
\newblock {\em JHEP}, 09:139, 2018, 1807.07915.

\bibitem{CMS:2018amk}
Albert~M Sirunyan et~al.
\newblock {Search for a new scalar resonance decaying to a pair of Z bosons in
  proton-proton collisions at $\sqrt{s}=13 $ TeV}.
\newblock {\em JHEP}, 06:127, 2018, 1804.01939.
\newblock [Erratum: JHEP 03, 128 (2019)].

\bibitem{CMS:2019qcx}
Albert~M Sirunyan et~al.
\newblock {Search for a heavy pseudoscalar boson decaying to a Z and a Higgs
  boson at $\sqrt{s} =$ 13 TeV}.
\newblock {\em Eur. Phys. J. C}, 79(7):564, 2019, 1903.00941.

\end{thebibliography}
%%%%%%%%%%%%%%%%%%%%%%%%%%%%%%%
\end{document}